\documentclass[aps,prb,amsmath,amssymb,showpacs,letterpaper,twocolumn]{revtex4}%
\usepackage{hyperref}
\usepackage{graphicx}
\usepackage{graphicx}%
\usepackage{amsfonts}%
\usepackage{color}
\usepackage{textcomp}

\begin{document}

\title{Multiscale modeling of a rectifying bipolar nanopore: Comparing Poisson-Nernst-Planck to Monte Carlo} 
\author{Bart\l{}omiej Matejczyk$^{1}$, M\'{o}nika Valisk\'{o}$^{2}$, Marie-Therese Wolfram$^{1,3}$, Jan-Frederik Pietschmann$^{4}$ and Dezs\H{o} Boda$^{2,5}$\footnote{Author for correspondence: boda@almos.vein.hu}}
\affiliation{$^{1}$Department of Mathematics, University of Warwick,
       CV4 7AL Coventry, United Kingdom}
\affiliation{$^{2}$Department of Physical Chemistry, University of Pannonia, P.O. Box 158, H-8201 Veszpr\'em, Hungary}
\affiliation{$^{3}$Radon Institute for Computational and Applied Mathematics, Austrian Academy of Sciences, Altenberger Strasse 69, 4040 Linz, Austria}
\affiliation{$^{4}$Institute for Computational and Applied Mathematics, WWU M\"unster, M\"unster 48149, Germany  }
\affiliation{$^{5}$Institute of Advanced Studies K\H{o}szeg (iASK), Chernel st. 14. H-9730 K\H{o}szeg, Hungary}
\date{\today}


\begin{abstract}
In the framework of a multiscale modeling approach, we present a systematic study of a bipolar rectifying nanopore using a continuum and a particle simulation method.
The common ground in the two methods  is the application of the Nernst-Planck (NP) equation to compute ion transport in the framework of the implicit-water electrolyte model.
The difference is that the Poisson-Boltzmann theory is used in the Poisson-Nernst-Planck (PNP) approach, while the Local Equilibrium Monte Carlo (LEMC) method is used in the particle simulation approach (NP+LEMC) to relate the concentration profile to the electrochemical potential profile.
Since we consider a bipolar pore which is short and narrow, we perform simulations using two-dimensional PNP. 
In addition, results of a non-linear version of PNP that takes crowding of ions into account are shown.
We observe that the mean field approximation applied in PNP is appropriate to reproduce the basic behavior of the bipolar nanopore (e.g., rectification) for varying parameters of the system (voltage, surface charge, electrolyte concentration, and pore radius). 
We present current data that characterize the nanopore's behavior as a device, as well as concentration, electrical potential, and electrochemical potential profiles.
\end{abstract}

\pacs{87.16.dp, 02.70.-c, 05.10.Ln, 07.05.Tp}

\maketitle



\section{Introduction}
\label{sec:intro}

In this paper, we compare Poisson-Nernst-Planck (PNP) theory with particle simulations for ionic transport through a rectifying bipolar nanopore. 
Both methods use the Nernst-Planck (NP) transport equation to describe the ionic flux of $i=\lbrace 1,2 \rbrace$ species:
\begin{equation}
 \mathbf{j}_{i}(\mathbf{r}) = -\dfrac{1}{kT}D_{i}(\mathbf{r})c_{i}(\mathbf{r})\nabla \mu_{i}(\mathbf{r}),
 \label{eq:np}
\end{equation} 
where $\mathbf{j}_{i}(\mathbf{r})$ is the particle flux density of ionic species $i$, $k$ the Boltzmann's constant, $T$ the temperature, and $D_{i}(\mathbf{r})$ the diffusion coefficient profile. 
The main difference between the two techniques is that PNP makes use of the Poisson-Boltzmann (PB) theory to relate the concentration profile, $c_{i}(\mathbf{r})$, to the electrochemical potential profile, $\mu_{i}(\mathbf{r})$, while the particle simulation method uses the Local Equilibrium Monte Carlo (LEMC) technique \cite{boda-jctc-8-824-2012,hato-jcp-137-054109-2012,boda-jml-189-100-2014,boda-arcc-2014} to establish this relation.
The particle simulation method includes all the ionic correlations that are beyond the mean field approximation applied in PNP. 
The difference between the two approaches can be quantified by considering the electrochemical potential
\begin{equation}
 \mu_{i}(\mathbf{r}) = \mu_{i}^{0} + kT\ln c_{i}(\mathbf{r}) + \mu_{i}^{\mathrm{EX}}(\mathbf{r}),
\label{eq:elchempot}
\end{equation} 
where $\mu_{i}^{0}$ is a standard chemical potential, a constant term that does not appear in the calculations.
The $\mu_{i}^{\mathrm{EX}}(\mathbf{r})$ term is the excess chemical potential that describes all the interactions acting between the particles forming the system and all the interactions with external forces (including an applied electrical potential).
PNP defines the excess term as the interaction with the mean electric field produced by all the free charges and induced charges.
Thus the electrochemical potential in the case of PNP is  
\begin{equation}
 \mu_{i}^{\mathrm{PNP}}(\mathbf{r}) = \mu_{i}^{0} + kT\ln c_{i}(\mathbf{r}) + z_{i}e\Phi(\mathbf{r}) ,
\label{eq:elchempotpnp}
\end{equation} 
where $z_{i}$ is the ionic valence, $e$ the elementary charge, and $\Phi(\mathbf{r})$ the total mean potential.
The missing term can be identified with what is beyond mean field (BMF) and quantifies the difference between PNP and a solution that is accurate from the point of view of statistical mechanics:
\begin{equation}
 \mu_{i}(\mathbf{r}) = \mu_{i}^{\mathrm{PNP}}(\mathbf{r}) +\mu_{i}^{\mathrm{BMF}}(\mathbf{r}) .
\label{eq:elchempotlemc}
\end{equation} 
In the implicit solvent framework used here the BMF term includes the volume exclusion effects (hard sphere effects) due to the finite size of the ions and electrostatic correlations that are beyond the mean-field level.
This partitioning has been used  to study selective adsorption of ions at electrodes \cite{valisko-jpcc-111-15575-2007} and in ion channels \cite{gillespie-bj-2008,boda-jcp-134-055102-2011,boda-cmp-18-13601-2015}.

It is also usual to break the electrochemical potential into a chemical and an electrical component that are loosely identified with the chemical and electrical works needed to bring an ion from one medium to the other:
\begin{equation}
\mu_{i}(\mathbf{r}) = \mu_{i}^{\mathrm{CH}}(\mathbf{r}) + \mu_{i}^{\mathrm{EL}}(\mathbf{r}) ,
\label{eq:ch-plus-el}
\end{equation} 
where the EL term can be identified with $z_{i}e\Phi(\mathbf{r})$, while the CH term can be identified with $\mu_{i}^{0}+kT\ln c_{i}(\mathbf{r}) + \mu_{i}^{\mathrm{BMF}}(\mathbf{r})$.
Although these two terms cannot be separated in experiments \cite{berry-rice-ross,bockris-reddy,fawcett-book}, the separation is possible in computational studies because $\Phi(\mathbf{r})$ can be determined. 
In PNP, where $ \mu_{i}^{\mathrm{BMF}}(\mathbf{r}) =0$, the CH term is just $\mu_{i}^{\mathrm{ID}}(\mathrm{r})=\mu^{0}_{i} + kT\ln c_{i}(\mathbf{r})$, the ideal expression (ID).

The PNP model is based on an approximate mean field approach with all the advantages and disadvantages.
First of all, the mean field method does not consider the particles as individual entities, but works with their concentration profiles which can be understood as the probability of finding an ion at a specific point in space and time. 
This probability depends on the interaction energy of the ion with the average (mean) electrical potential produced by all the charges in the system, including all the ions.
Two- and many-body correlations between ions, therefore, are neglected in PNP. 
The ions are treated as point charges omitting their size. 

In this work, we also use a non-linear variant of PNP (denoted by nPNP from now on) that can be derived (formally) from a discrete hopping model\cite{burger2010nonlinear}.
Similar models have been derived by Bikerman\cite{Bikerman1942} and Li\cite{Li2009}.
In all these models Eqs.\ \eqref{eq:np} and \eqref{eq:elchempotpnp} are replaced by
\begin{equation}\label{e:nnp}
 \mathbf{j}^{\mathrm{nPNP}}_i(\mathbf{r}) = -\frac{1}{kT} D_i(\mathbf{r}) c_i (\mathbf{r}) \alpha (\mathbf{r}) \nabla \mu_{i}^{\mathrm{nPNP}}(\mathbf{r})
\end{equation}
and
\begin{equation}\label{e:munnp}
 \mu_{i}^{\mathrm{nPNP}}(\mathbf{r}) = \mu_{i}^{\mathrm{PNP}}(\mathbf{r})  - kT \ln \alpha (\mathbf{r}) ,
\end{equation} 
where
\begin{equation}
\alpha(\mathbf{r}) = 1- \frac{c_1(\mathbf{r})} {c_{\max}}-\frac{c_2(\mathbf{r})}{c_{\max}} ,
\end{equation} 
and we have chosen $c_{\max} = 61.5$ mol/dm$^{3}$ as a maximum value for the concentation at close packing.
The scaling factor, $\alpha(\mathbf{r})$ approaches 1 as $c_{i}\rightarrow 0$, so nPNP turns into PNP in this limit.
Another approach to overcome these limitation is density functional theory (DFT) \cite{rosenfeld-prl-63-980-1989,rosenfeld-jcp-98-8126-1993,kierlik-pra-42-3382-1990,yteran_jcp_1990,kierlik-pra-44-5025-1991,rosenfeld-pre-55-4245-1997,groh-pre-57-6944-1998,gillespie_jpcm_2002,gillespie_pre_2003} which includes additional terms in the free energy that take care of the interactions in the BMF term. 
A recent review discusses different possibilities to account for ions size \cite{gillespie_mfnf_2015}.

In most of the literature, a one-dimensional (1D) reduction of PNP is used. 
This gives good results especially for long and narrow nanopores \cite{vlassiouk_acsnanno_2008} as its derivation is based on the assumption that the radius is significantly smaller than the length.  
Its advantage is that it requires less computational effort and makes the computation of long pores possible.  
In this work, however, we use a two dimensional (2D) PNP (respectively nPNP) model which is a suitable approximation to the three-dimensional (3D), but rotationally symmetric, system studied here.
Our model, furthermore, includes the  bulk regions and the access regions at the entrances of the nanopore, as opposed to other studies  \cite{vlassiouk_acsnanno_2008, vlassiouk_jacs_2009}.
Solving PNP in a larger domain and also in the radial dimension gives more accurate results.

Summarised, we can couple the NP equation either to LEMC simulations or to the PB theory.
The former is referred to as the NP+LEMC technique, while the latter could be termed as NP+PB, but we stay with the usual name, PNP.
Poisson's equation is satisfied in both methods.
In PNP, it is solved in every iteration, while it is automatically fulfilled in LEMC because Coulomb's law is used to handle electrostatics in the simulations (including the applied field in the framework of the Induced Charge Computation method \cite{boda-pre-69-046702-2004,boda-jcp-125-034901-2006}).
Both approaches provide approximate indirect solutions for the dynamical problem through the NP equation.
Direct simulation of ionic transport in the implicit solvent framework is commonly done by Brownian Dynamics (BD) simulations \cite{chung-bj-77-2517-1999,im_bj_2000,berti-jctc-10-2911-2014}.
The main difference between NP+LEMC and PNP is the way they handle the statistical mechanical problem of establishing the closure between $c_{i}(\mathbf{r})$ and $\mu_{i}(\mathbf{r})$. 
The NP+LEMC technique provides a solution on the basis of particle simulations that contain all the correlations ignored by PNP.
The main goal of our study is to discuss the effects of the approximations applied in PNP for different sets of physical parameters.
Comparing to NP+LEMC results makes it possible to focus on the approximations applied in the statistical mechanical part of the PNP theory (the PB theory), because NP is common in them.
If we want to reveal the magnitude and nature of errors resulting from the application of the approximative NP equation instead of simulating ion transport directly, we need to compare to BD  \cite{moy-bj-2000,corry-bj-2000}.
Comparison between BD and NP+LEMC results will be published in a separate paper.

We apply our methods to a bipolar nanopore that is a suitable case study for our purpose. 
Bipolar nanopores have an asymmetrical surface charge distribution on the pore wall changing sign along the central axis of the pore.
Pore regions with opposite surface charges can be achieved by chemical modifications. 
For example, in the case of PET nanopores, carboxyl groups can be transformed into amino groups by a coupling agent \cite{vlassiouk_nl_2007}.
The surface potential can also be regulated similarly to field-effect transistors if the pore walls are made of conducting materials.

The reason of choosing the bipolar nanopore for this comparative work is that the source of rectification in this case is purely electrostatic in nature and thus a robust effect. 
Therefore, we can afford a short nanopore (only 6 nm in length) that can be handled with LEMC. 
In the case of conical nanopores, where only a geometrical asymmetry is present, long pores are needed to produce a considerable effect which makes it computationally unfeasible. 

Although bipolar nanopores have been studied extensively using PNP \cite{daiguji_nl_2005,constantin_pre_2007,vlassiouk_nl_2007,karnik_nl_2007,vlassiouk_acsnanno_2008,kalman_am_2008,yan_nl_2009,nguyen_nt_2010,szymczyk_jpcb_2010,singh_jap_2011,singh_jpcb_2011,singh_apl_2011,van_oeffelen_plosone_2015,tajparast_bba_2015}, we are not aware of any paper, where a direct comparison to particle simulations is discussed for this system.
Furthermore, most of those works use 1D PNP, while we report results of 2D PNP here and the comparison with nPNP is also completely new. 

Particle simulations are necessary for narrow pores, where ions are crowded and their size and the correlations between them (the BMF term) matter.
This is the case in ion channels, where the ions correlate strongly with each other and with the charged amino acids along the ionic pathway.
Although nanopores are larger in reality, the electrical double layers formed by the ions at the pore walls overlap if the the Debye length is larger than the pore radius.
This occurs if the pore is narrow enough (such as conical nanopores at their tips) or if the electrolyte is dilute. 

This work belongs to a series of studies, where we apply a multiscale modeling approach \cite{steinhauser_multiscale_2008}.
We can create different models (with less or more details) and we can study these models with computational methods that fit the model.
In another recent work, \cite{hato_nl_2017}, we compared results of molecular dynamics (MD) simulations performed for an all-atom model including explicit water to NP+LEMC calculations performed for the implicit-water model (the same model studied in this paper). 
We concluded that, despite all the simplifications, the implicit-water model provides an appropriate framework to study nanopores. 
The link between the two modeling levels is the diffusion coefficient profile, $D_{i}(\mathbf{r})$, used in NP+LEMC as an input, while the MD simulations can provide information about this profile.

The advantage of NP+LEMC over MD is that it is faster and can handle larger systems that are closer to realistic length scales of nanodevices. 
From this point of view, PNP is even more advantageous, because it does not involve particle simulations, therefore, it can handle even larger systems. 

\section{Models and methods}
\label{sec:modelsandmethods}

\subsection{Models}
\label{sec:models}

When we extract macroscopic information (currents, profiles, etc.) from a microscopic model, we construct a model that contains the interactions between the particles and the external constraints (hard walls, applied field, etc.).
This is equivalent with defining the Hamiltonian of the system precisely.
This model then can be studied with different statistical mechanical methods (simulations or theories).
Whether a disagreement with experiments is due to oversimplifications in the modeling or the approximations in the method can be sorted out by comparing to particle simulations, where approximations in the method are usually absent (system size errors and statistical noises are still present).

Although the separation of model and method is not so distinct in PNP, we describe the two models together in this subsection in order to emphasize similarities and differences.
Note, however, that the term ``method'' in the case of PNP refers to the physical equations used in PNP (see next subsection), not the numerical method with which we solve the PNP equation.

The electrolyte is modeled in the implicit solvent framework in both cases. 
Water is a continuum background, whose energetic effect is taken into account by a dielectric screening ($\epsilon = 78.5$ in the denominator of the Coulomb potential and as coefficient in the Poisson equation, respectively), while its dynamic effect is included in the diffusion coefficient in the NP equation ($D_{i}(\mathbf{r})$ in Eq.\ \ref{eq:np}). 
The ions are point charges in PNP, while they are hard spheres (of radius 0.3 nm for both ions) with point charges at their centers in LEMC.

\begin{figure*}[t]
 \begin{center}
\scalebox{0.34}{\includegraphics*{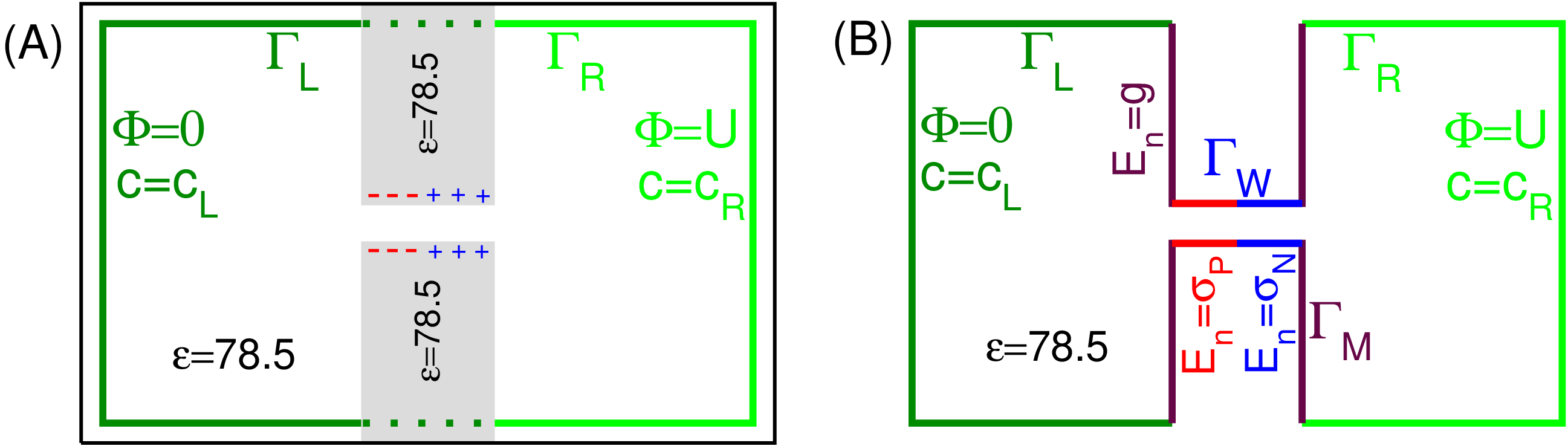}}
 \end{center}
\caption{\small Geometry of computation domain (A) in the NP+LEMC system and (B) in the PNP system.
(A) Boundary conditions for the NP+LEMC system are prescribed for the two half-cylinders (dark and light green lines, $\Gamma_{\mathrm{L}}$ and $\Gamma_{\mathrm{R}}$ domains) on the two sides of the membrane. 
Dirichlet boundary conditions are applied by using the appropriate applied potential obtained by solving the Laplace equation (a linear interpolation way used inside the membrane, see the dotted lines). 
Boundary conditions for the concentrations are ensured by using appropriate electrochemical potentials at the boundaries that correspond to the chemical potentials producing the prescribed concentrations. 
The domains outside the green lines are in thermodynamic equilibrium, where the chemical potential is constant, so equilibrium GCMC simulations are performed there.
Pore charges are free charges present explicitly in the simulation cell. 
They are placed on the pore wall on a grid as partial point charges.
The dielectric constant is the same everywhere, including the interior of the membrane.
(B) The PNP computational cell excludes the interior of the membrane from the solution domain.
The pore charges are polarization charges that are induced as a result of the prescribed Neumann boundary conditions on the pore wall (red and blue lines, $\Gamma_{\mathrm{W}}$).
On the surface of the membrane (brown lines, $\Gamma_{\mathrm{M}}$), a Neumann boundary conditions is applied in order to mimic the NP+LEMC solution.
On the two half cylinders, the same boundary conditions are used as in NP+LEMC ($\Gamma_{\mathrm{L}}$ and $\Gamma_{\mathrm{R}}$).
}
\label{Fig1}
\end{figure*}

The nanopore is a cylinder of 6 nm in length with a varying radius ($R=0.5-3$ nm). 
It penetrates a membrane that separates two bulk electrolytes.
The walls of the pore and the membrane are hard impenetrable surfaces in the LEMC simulations (Fig.\ \ref{Fig1}A), while they are part of the boundaries of the solution domain in the PNP calculations (Fig.\ \ref{Fig1}B).

The diffusion coefficient of the ions are usually smaller inside the pore than outside in the bulk regions.
This finding was confirmed by our other study that compares MD and NP+LEMC results \cite{hato_nl_2017}.
Here, for simplicity, we assigned $D_{i}^{\mathrm{bulk}}=1.333\times 10^{-9}$ m$^{2}$s$^{-1}$ and $D_{i}^{\mathrm{pore}}=1.333\times 10^{-10}$ m$^{2}$s$^{-1}$ values in the bulk and in the pore, respectively, for both ions.

The charges on the cylinder's surface are partial point charges in the case of LEMC that are placed on grid points whose average distance is about 0.25 nm.
The values of the partial charges depend on the prescribed surface charge density, $\sigma$.
The surface charge densities are included in PNP through Neumann boundary conditions for the potential.

Athough ions are excluded from the interior of the membrane by hard walls in LEMC, the electric field is still present there and could be computed. 
The dielectric constant is the same there as in the electrolyte ($\epsilon = 78.5$), therefore, the surface of the membrane is not a dielectric boundary and polarization charges are not induced there  (Fig.\ \ref{Fig1}A).
In the case of PNP, the interior of the membrane is not part of the computational domain (Fig.\ \ref{Fig1}B).
An appropriate Neumann boundary condition is applied on the surface of the membrane in order to mimic the system used in LEMC.
Boundary conditions as handled in the two methods are detailed in the following subsections.

\subsection{Poisson-Nernst-Planck theory}
\label{sec:pnp}

Introduced for modeling semiconductors, \cite{markowich1990semiconductor,Mar}, PNP was soon adapted to the modeling of biological ion channels \cite{chen_bj_1992,eisenberg_jmb_1996,chen_bj_1997,chen-bj-73-1337-1997,chen_siam_1997,nonner_bj74_1998,nonner_bj75_1998,chen_bj_1999,moy-bj-2000,corry-bj-2000,nonner_bj_2000,richardson2009multiscale} as well as synthetic nanopores \cite{pietschmann_pccp_2013,daiguji2004ion,siwy_prl_2002}. 
The classical PNP is a self-consistent system providing a flux that satisfies the continuity equation 
\begin{equation}\label{eq:continuity}
\nabla \cdot \mathbf{j}_{i}(\mathbf{r})=0.
\end{equation} 
The flux is computed from the NP equation (Eq.\ \ref{eq:np}) for the linear and Eq. \ref{e:nnp} for the nonlinear version) with the electrochemical potential, $\mu_i$, defined in Eqs.\ \ref{eq:elchempotpnp} and \ref{e:munnp}, respectively. 
The mean electrostatic potential, $\Phi(\mathbf{r})$, is connected to the concentration profiles in both cases through the Poisson equation:
\begin{equation}
 -\nabla \cdot (\varepsilon \nabla \Phi) = \dfrac{e}{\epsilon_0} \sum_{i}z_i n_i(\mathbf{r}).
\end{equation} 
Here $n_{i}(\mathbf{r})$ is the number density of ions (measured in m$^{-3}$) connected to concentration (measured in mol/dm$^{3}$) through $n_{i}(\mathbf{r})=1000N_{\mathrm{A}}c_{i}(\mathbf{r})$.
We further denote by $N_{\mathrm{A}}$ Avogadro's number and by $\epsilon_0$ the permittivity of vacuum.
 
Equation \ref{eq:elchempotpnp} is equivalent to the application of Boltzmann's distribution, which provides the statistical mechanical description of the system.
Together with the Poisson equation it forms the PB theory.
Coupling these to the NP equation applies the theory for a non-equlibrium situation and provides the PNP theory.
From a physical chemical point of view, Eq.\ \ref{eq:elchempotpnp} corresponds to the statement that the electrolyte solution is ideal ($\mu_{i}^{\mathrm{BMF}}(\mathbf{r})=0$).
While many theories were developed in the last years to overcome this limitation (e.g. DFT), here we employ a variant of PNP that features non-linear cross-diffusion in the continuity equations which limits the maximal density and thus takes into account the finite size of the ions.  

The (n)PNP systems are solved inside the computational domain, whose boundary is separated into four parts as shown in Fig.\ \ref{Fig1}B. 
The first two parts correspond to the left and right half-cylinders (dark and light green lines in Fig.\ \ref{Fig1}B) and are denoted by $\Gamma_\mathrm{L}$ and $\Gamma_\mathrm{R}$. 
These regions are the same in NP+LEMC.
Both the concentration and the potential are set using the following boundary conditions
\begin{align}
c_i (\mathbf{r}) = c_{i}^{\mathrm{L}} & \quad \text{and} \quad \Phi(\mathbf{r})=0 \quad \text{on } \Gamma_\mathrm{L} \nonumber \\
c_i (\mathbf{r}) = c_{i}^{\mathrm{R}} & \quad \text{and} \quad \Phi(\mathbf{r})=U \quad \text{on } \Gamma_\mathrm{R} 
\label{eq:bc-pnp1}
\end{align}
The third part are the regions of the membrane which are attached to the baths and are denoted by $\Gamma_\mathrm{M}$ (brown lines in Fig.\ \ref{Fig1}B). 
As the membrane is impenetrable for the particle flux, we set the flux to be equal to 0 there.  
In LEMC  simulations the membrane is penetrable for the electric field, which is not the case in PNP. 
Therefore we impose the boundary conditions
\begin{align}\label{eq:additionalneumann}
 \mathbf{j}_{i}(\mathbf{r})\cdot \mathbf{n}_{\mathrm{M}}=0  & \quad \text{and} \quad \frac{\partial \Phi (\mathbf{r})}{\partial \mathbf{n}_{\mathrm{M}}}  = g(r) \quad \text{on }\Gamma_\mathrm{M},
\end{align}
where $\mathbf{n}_{\mathrm{M}}$ is the outer normal on $\Gamma_\mathrm{M}$ and the function $g(r)$ is supposed to mimic the LEMC case (where there is an electric field across the membrane).
More precisely, it is obtained by solving a Laplace equation with zero left hand side without permanent charges and with boundary condition Eq. \ref{eq:bc-pnp1} in the domain of Fig.\ \ref{Fig1}A. 
Then, evaluating the normal derivative of this solution at the boundary $\Gamma_\mathrm{M}$ yields the function $g(\mathbf{r})$.
This additional Neumann boundary condition matches the value of applied potential crossing the membrane in the LEMC.

The last part of the boundary is on the inside wall of the pore, called $\Gamma_\mathrm{W}$. 
As it is a part of the membrane, which is impenetrable for the particles, no-flux conditions are also imposed for the current. 
The permanent charges induce an additional electric field and are included by another Neumann boundary condition: 
\begin{align}
 \mathbf{j}_{i}(\mathbf{r})\cdot \mathbf{n}_{\mathrm{W}}=0  & \quad \text{and} \quad \frac{\partial \Phi (\mathbf{r})}{\partial \mathbf{n}_{\mathrm{W}}}  = \sigma_{0}(z) \quad \text{on } \Gamma_\mathrm{W},
\end{align}
where $\sigma_{0}= \sigma$ and $\sigma_{0}= - \sigma$  for $z<0$ and $z>0$, respectively, and $\mathbf{n}_{\mathrm{W}}$ is the outer normal on $\Gamma_\mathrm{W}$.

One of the most popular simplification of the full PNP model is the 1D reduction. 
It is broadly used and compared with experimental data \cite{pietschmann_pccp_2013}. 
It also allows to simulate very long nanopores with  complex geometry using reasonable computational time. 
The derivation of the 1D model is based on the assumption that the length of the nanopore is significantly larger than its radius. 
Since in our setup this is not the case, we perform all simulations in two spatial dimensions.
 
To actually solve the 2D (n)PNP system we use the well-known Scharfetter--Gummel scheme which is based on a transformed formulation of the system in exponential variables, see \cite{gummel1964self} for detail. 
We use a 2D finite element method for the actual implementation and a triangular mesh containing $20-60$ thousand elements, depending on the radius of the pore. 
The mesh is also non-uniform in order to obtain high accuracy, especially close to the pore entrances.

\subsection{Nernst-Plank equation coupled to Local Equilibrium Monte Carlo }
\label{sec:nplemc}

To solve the NP+LEMC system, an iterative procedure is needed, where $\mu_{i}$ is updated until the continuity equation (Eq.\ \ref{eq:continuity}) is satisfied.
The procedure can be summarized as
\begin{equation}
\mu_{i}[n] \,\,\, \xrightarrow{\mathrm{LEMC}}  \,\,\,  c_{i}[n] \,\,\,  \xrightarrow{\mathrm{NP}} \,\,\,  \mathbf{j}_{i}[n] \,\,\, 
\xrightarrow{\nabla \cdot \mathbf{j}=0} 
\,\,\, \mu_{i}[n+1] .
\label{eq:circle}
\end{equation} 
The electrochemical potential for the next iteration, $\mu_{i}[n+1]$, is computed from the results of the previous iteration, $c_{i}[n]$, in a way that they together produce a flux (through the NP equation) that satisfies the continuity equation.
Details on the algorithm can be found in our original paper \cite{boda-jctc-8-824-2012}.

The concentration profile in an iteration, $c_{i}[n]$, corresponding to the electrochemical potential profile, $\mu_{i}[n]$, is obtained from LEMC simulations.
We divide the computational domain (inside the green lines in Fig.\  \ref{Fig1}A) into volume elements and assume local equilibrium in these volume elements.
We assume that these local equilibria can be characterized by local electrochemical potential values.
We also assume that the gradient of the $\mu_{i}$ profile defined this way is the driving force of ion transport as described by the NP equation (Eq.\ \ref{eq:np}).  

The heart of the LEMC simulation is a MC step, where we insert/remove an ion into/from a volume element $\mathcal{B}^{k}$. 
The acceptance probability of an insertion is 
\begin{equation}
 p_{i,\mathrm{INS}}^{k} = 
\min\left\lbrace 1, \dfrac{V^{k}}{N_{i}^{k}+1} \exp \left( \dfrac{-\Delta U^{k} + \mu_{i}^{k}}{kT} \right) \right\rbrace ,
\label{eq:gcmc-pin}
\end{equation} 
where $V^{k}$ is the volume of $\mathcal{B}^{k}$, $N_{i}^{k}$ is the number of particles of component $i$ in $\mathcal{B}^{k}$ before insertion, $\Delta U^{k}$ is the energy change associated with the insertion (including the effect of the external field), and $\mu_{i}^{k}$ is the configurational (total minus $\mu_{i}^{0}$) electrochemical potential of component $i$ in $\mathcal{B}^{k}$.
In the particle deletion step we randomly choose a particle of component $i$ in sub-volume $\mathcal{B}^{k}$ and delete it. 
The deletion is accepted with probability
\begin{equation}
 p^{k}_{i,\mathrm{DEL}} =\min\left\lbrace 1, \dfrac{N^{k}_{i}}{V^{k}} \exp \left( \dfrac{-\Delta U^{k} - \mu_{i}^{k}}{kT} \right) \right\rbrace .
\label{eq:gcmc-pout}
\end{equation} 
Here, $N_{i}^{k}$ is the number of particles of component $i$ in subvolume $\mathcal{B}^{k}$ before deletion.

The energy change $\Delta U^{k}$ contains the effect of the full simulation domain outside subvolume $\mathcal{B}^{k}$ including short-range interactions such hard-sphere exclusions between ions and hard-wall exclusion with membrane wall.
The configurational space is sampled properly, because the ions experience the potential produced by billions of possible configurations, not just a mean potential as in the case of PNP. 

The effect of the applied potential is also included in $\Delta U^{k}$.
The applied potential is computed by solving the Laplace equation with the Dirichlet boundary condition of Eq.\ \ref{eq:bc-pnp1} for the boundary surface confining the solution domain.
The boundary conditions for concentrations (see Eq.\ \ref{eq:bc-pnp1}) are set by finding the appropriate chemical potentials in the two baths that produce the desired concentrations in the GCMC simulations.
We used the Adaptive GCMC method \cite{malasics-jcp-132-244103-2010} to determine these chemical potentials. 

The result of the simulation is the concentration $c_{i}^{k}$ in every volume element.
The values $c_{i}^{k}$ and $\mu_{i}^{k}$ are assigned  to the centers of the volume elements and so the corresponding profiles are constructed.
Both $c_{i}^{k}$ and $\mu_{i}^{k}$ fluctuate during the iteration process, so the final results are obtained as running averages.

The NP+LEMC technique has been applied to study transport through membranes \cite{boda-jctc-8-824-2012,hato-jcp-137-054109-2012}, calcium channels \cite{boda-jml-189-100-2014,boda-arcc-2014}, and bipolar nanopores \cite{hato_nl_2017}.

\section{Results and Discussion}\label{sec:results}

The reference point of all simulations corresponds to the following parameter set: voltages $\pm 200$ mV (200 mV is the ON, while -200 mV is the OFF state of the nanopore), concentrations $c=0.1$ and 1 M, surface charge $\sigma =1$ $e$/nm$^{2}$, and nanopore radius $R=1$ nm.
Then, we vary the parameters systematically by changing only one and keeping the others fixed. Rectification is defined by $|I(U)/I(-U)|$, i.e. the ratio between the currents in the ON and OFF state, respectively. 
In our case, this implies that it is always larger than $1$.
In all figures we plot the NP+LEMC, PNP, and nPNP results with symbols, solid lines, and dashed lines, respectively.

\begin{figure}[t]
 \begin{center}
\scalebox{0.5}{\includegraphics*{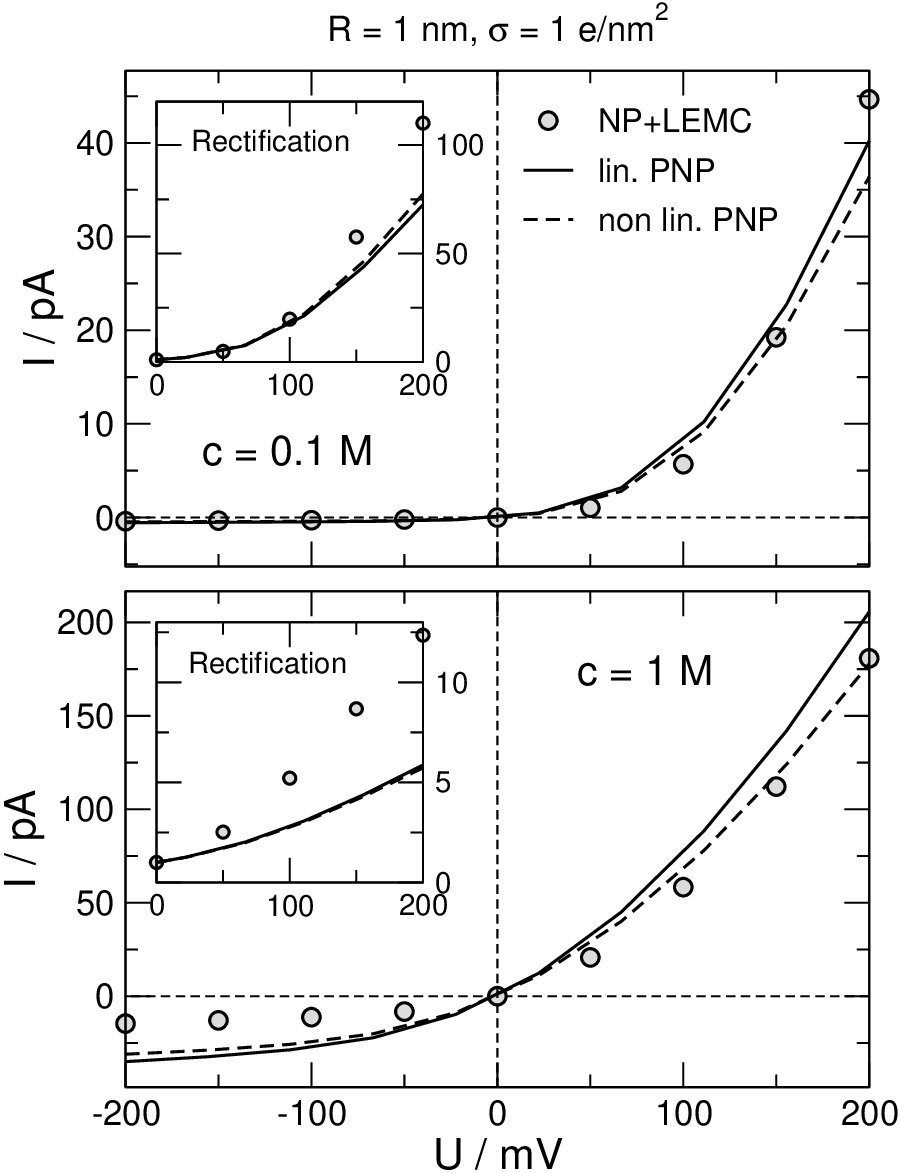}}
 \end{center}
\caption{\small Current-voltage curves for concentrations $c=0.1$ M (top panel) and $c=1$ M (bottom panel) as obtained from NP+LEMC, PNP, and nPNP (symbols, solid curves, dashed curves, respectively).
The insets show rectification as computed from the ratio of the ON and OFF state currents (the absolute values).
The model parameters are $R=1$ nm and $\sigma =1$ $e$/nm$^{2}$.
}
\label{Fig2}
\end{figure} 

\subsection{Comparison of I-U curves and rectification behavior}

First, we look at the nanopore as a device that gives an output signal (current) as an answer to the input signal (voltage).
The relation of these is the transfer function of the device.
Then, we study various profiles (concentration, potential, chemical potential) and try to understand the differences between PNP and NP+LEMC.

Figure \ref{Fig2} shows current-voltage (I-U) curves for the concentrations $c=0.1$ and 1 M.
Rectification is observed using all the three methods: the current is larger at positive voltages than at negative voltages (note that electrical currents are multiplied with -1 in order to get positive currents for positive voltages).
Rectification increases with increasing $|U|$ as shown in the insets.
Agreement between NP+LEMC and (n)PNP data is better at low concentration (0.1 M) and smaller voltages as expected.
The data from nPNP are slightly better than those from PNP, especially for $c=1$ M.

\begin{figure}[t]
 \begin{center}
\scalebox{0.5}{\includegraphics*{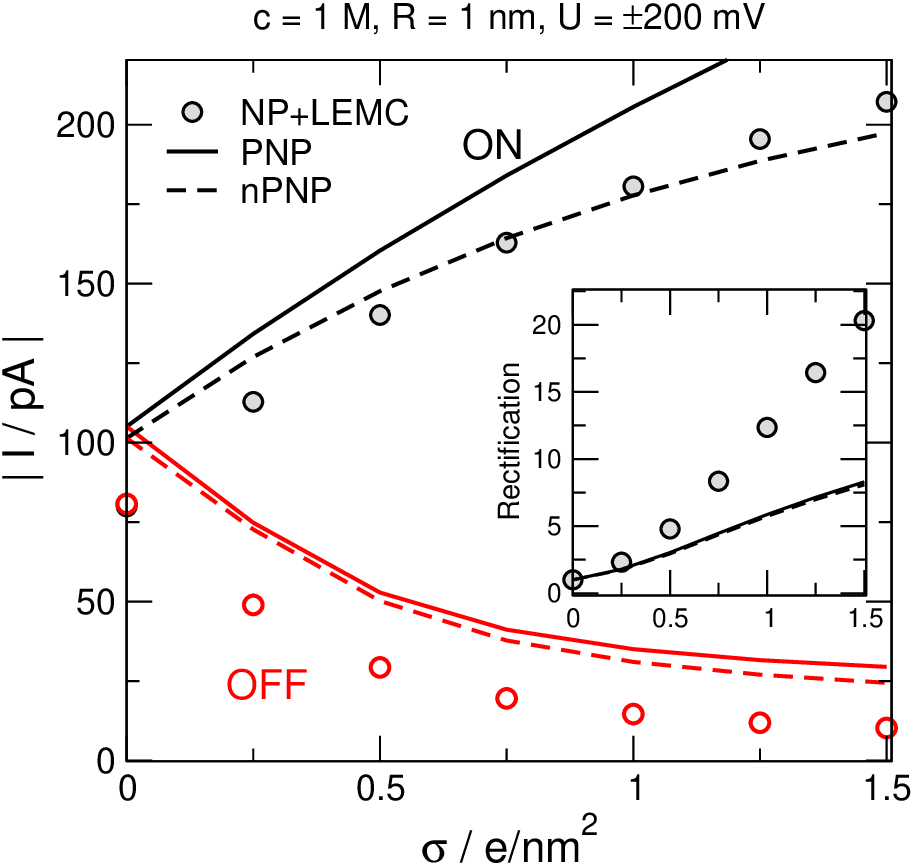}}
 \end{center}
\caption{\small The absolute value of the current as a function of $\sigma$ (characterizing the strength of the polarity of the pore) in the ON and OFF states (200 vs.\ -200 mV, respectively) as obtained from NP+LEMC, PNP, and nPNP (symbols, solid curves, dashed curves, respectively).
The inset shows rectification.
The model parameters are $c=1$ M and $R=1$ nm.
}
\label{Fig3}
\end{figure}

The value of the $\sigma$ parameter can be considered as a measure of the nanopore's polarity.
At $\sigma=0$ $e$/nm$^{2}$, the pore is uncharged and symmetric, so currents at the two voltages of opposite signs are the same and rectification is 1.
Figure \ref{Fig3} shows current values in the ON and OFF states as functions of $\sigma$.
As $\sigma$ is increased, the current increases in the ON state, while decreases in the OFF state.
Rectification, therefore, improves as the strength of the polarity of the pore increases.
The $\sigma$-dependence is well described by (n)PNP qualitatively.
The errors manifest in the fact that rectification is underestimated by (n)PNP.

One source of the errors is that the effective cross section of the pore through which the centers of ions can move is smaller in the case of the charged hard sphere ions used in LEMC ($R-0.15$ nm, where 0.15 nm is the ionic radius) than in the case of point ions used in (n)PNP (the whole pore radius, $R$, is used in (n)PNP).
(n)PNP, therefore, systematically overestimates current in both the ON and OFF states as seen in Fig.\ \ref{Fig3}.
The overestimation of the denominator (OFF current) dominates the ratio.
Rectification, therefore, is underestimated. 

One way to partially overcome this difference between the two models would be using the effective cross section of the finite ions ($R-0.15$ nm) in the PNP calculations. 
In this case, Fig.\ \ref{Fig3} would show better agreement, but cause other problems, such as the presence of ions with different diameters.
Therefore, we decided to keep the pore cross section in PNP in this study as it is ($R$), but point out the problems with this approach.

\begin{figure}[t]
 \begin{center}
\scalebox{0.5}{\includegraphics*{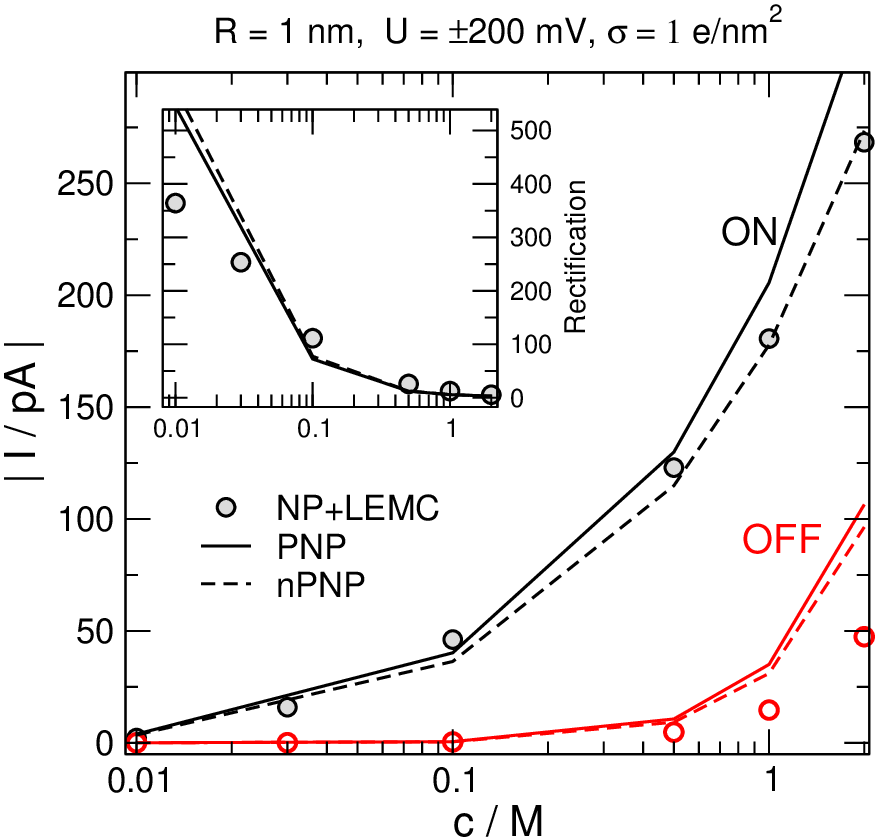}}
 \end{center}
\caption{\small The absolute value of the current as a function of the electrolyte concentration in the ON and OFF states ($200$ vs.\ $-200$ mV, respectively) as obtained from NP+LEMC, PNP, and nPNP (symbols, solid curves, dashed curves, respectively).
The inset shows rectification.
The model parameters are $c=1$ M and $R=1$ nm. 
}
\label{Fig4}
\end{figure}

Figure \ref{Fig4} shows the currents as functions of the electrolyte concentration, $c$.
Currents decrease with decreasing $c$ as expected, but the current decreases faster in the OFF state, so rectification increases with decreasing concentration, a well-known result.
The explanation is that depletion zones dominate the currents in bipolar nanopores, but depletion zones are more depleted at low concentrations.
Changing the sign of the voltage from positive (ON) to negative (OFF), therefore, can deplete the depletion zone further more efficiently at low concentrations.

Agreement between NP+LEMC and (n)PNP is better in the ON state.
The nonlinear version of PNP works better in this case, because it handles crowding better.
In the OFF state, (n)PNP systematically overestimates the current partly from the reason discussed above. 
Rectification, interestingly, is underestimated by (n)PNP at large, while overestimated at small concentrations.

\begin{figure}[t]
 \begin{center}
\scalebox{0.5}{\includegraphics*{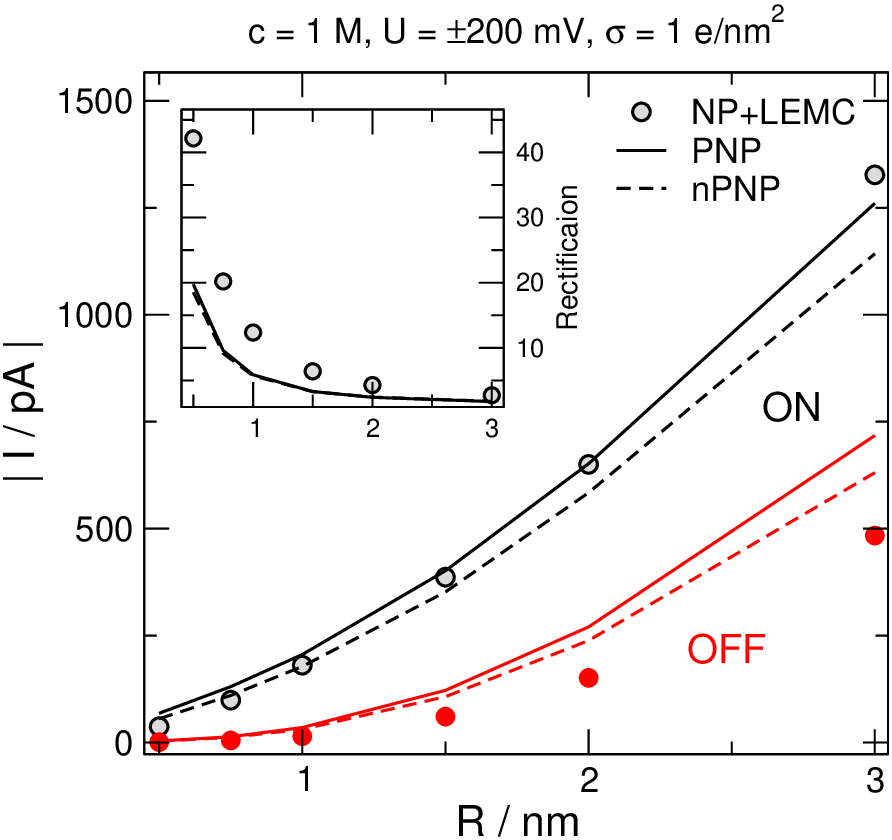}}
 \end{center}
\caption{\small The absolute value of the current as a function of the pore radius in the ON and OFF states (200 vs.\ -200 mV, respectively) as obtained from NP+LEMC, PNP, and nPNP (symbols, solid curves, dashed curves, respectively).
The inset shows rectification.
The model parameters are $R=1$ nm and $\sigma=1$ $e$/nm$^{2}$.
}
\label{Fig5}
\end{figure}

Finally, we show the dependence of currents on the pore radius in Fig.\ \ref{Fig5}.
Currents increase with widening pores as expected.
The relative difference between the ON and OFF states decreases as $R$ increases.
Rectification is the result of the interplay between the effect of pore charges and the applied potential. 
The average distance of pore charges from the ions increases as $R$ increases, therefore, the pore charges get less and less able to produce the depletion zones inside the pore.
(n)PNP qualitatively reproduces the behavior obtained from NP+LEMC.
Also, the systematic underestimation of rectification is present for all pore radii studied.

\subsection{Analysis of profiles for concentration, electrical potential, and electrochemical potential}

To get additional insights into the physical mechanisms beyond the device-level behavior, we also analyze profiles for the concentration, electrical potential, and electrochemical potential.

\begin{figure*}[t]
 \begin{center}
\scalebox{0.48}{\includegraphics*{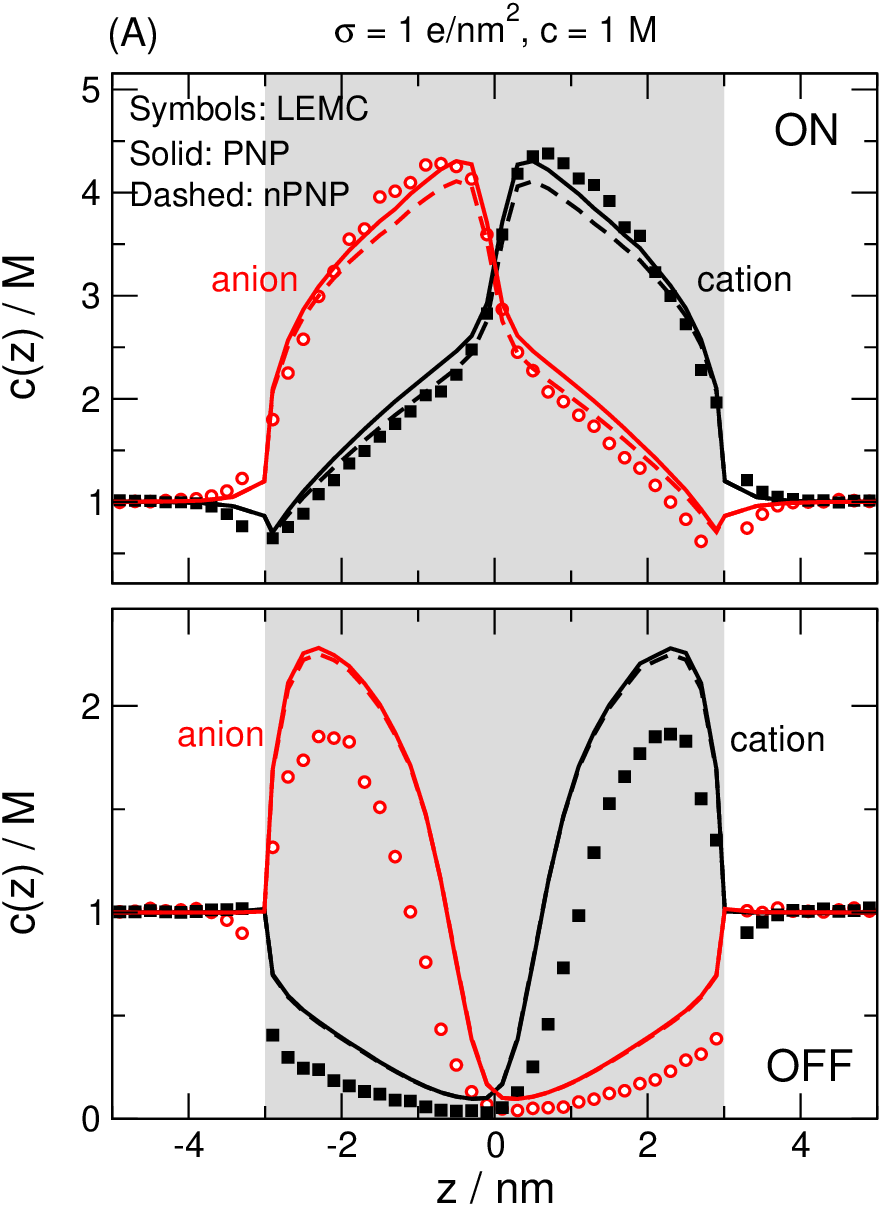}}\scalebox{0.4934}{\includegraphics*{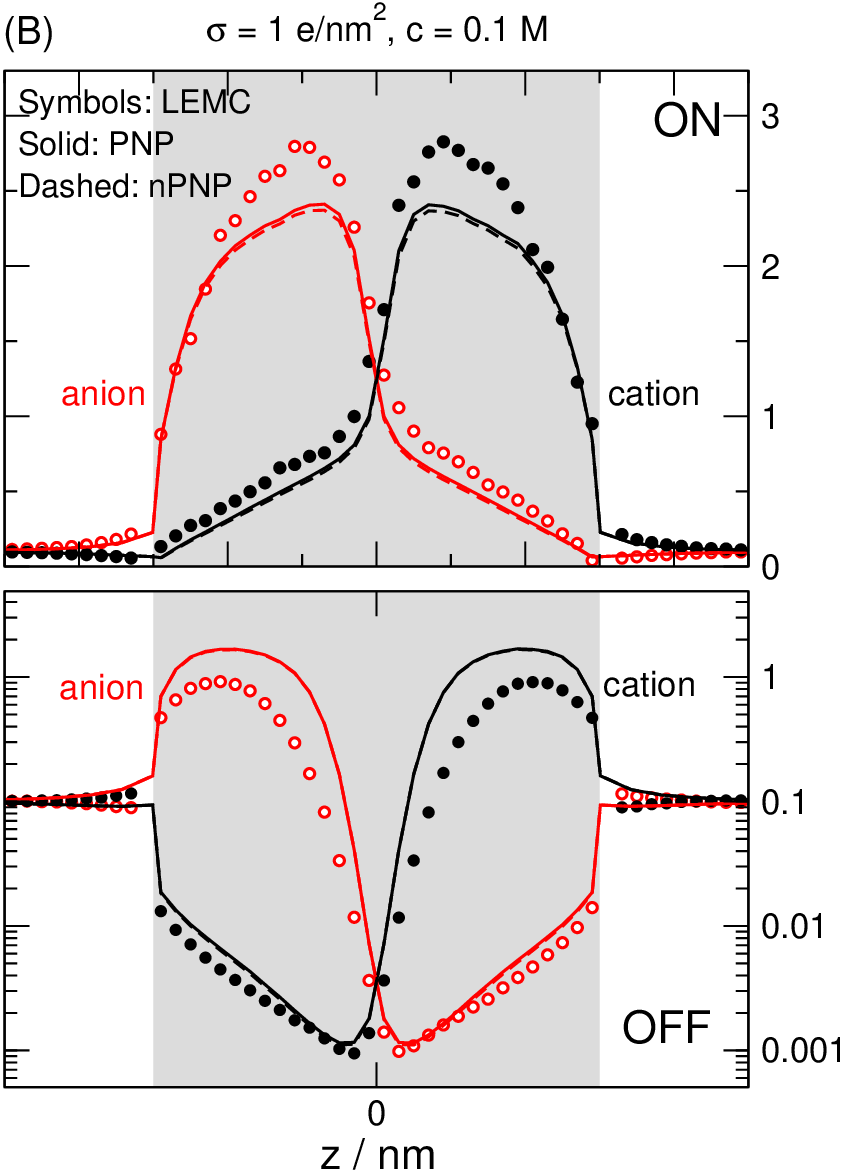}}
 \end{center}
\caption{\small 
Concentration profiles of cations and anions as obtained from NP+LEMC, PNP, and nPNP for (A) $c=1$ M and  (B) $c=0.1$ M for parameters $R=1$ and $\sigma=1$ $e$/nm$^{2}$.
These concentration profiles have been computed by taking the average number of ions in a slab and dividing by the available volume.
For $-3<z<3$ nm, the cross section of the pore was used to obtain this volume in both methods.
}
\label{Fig6}
\end{figure*}

\begin{figure*}[t]
 \begin{center}
\scalebox{0.48}{\includegraphics*{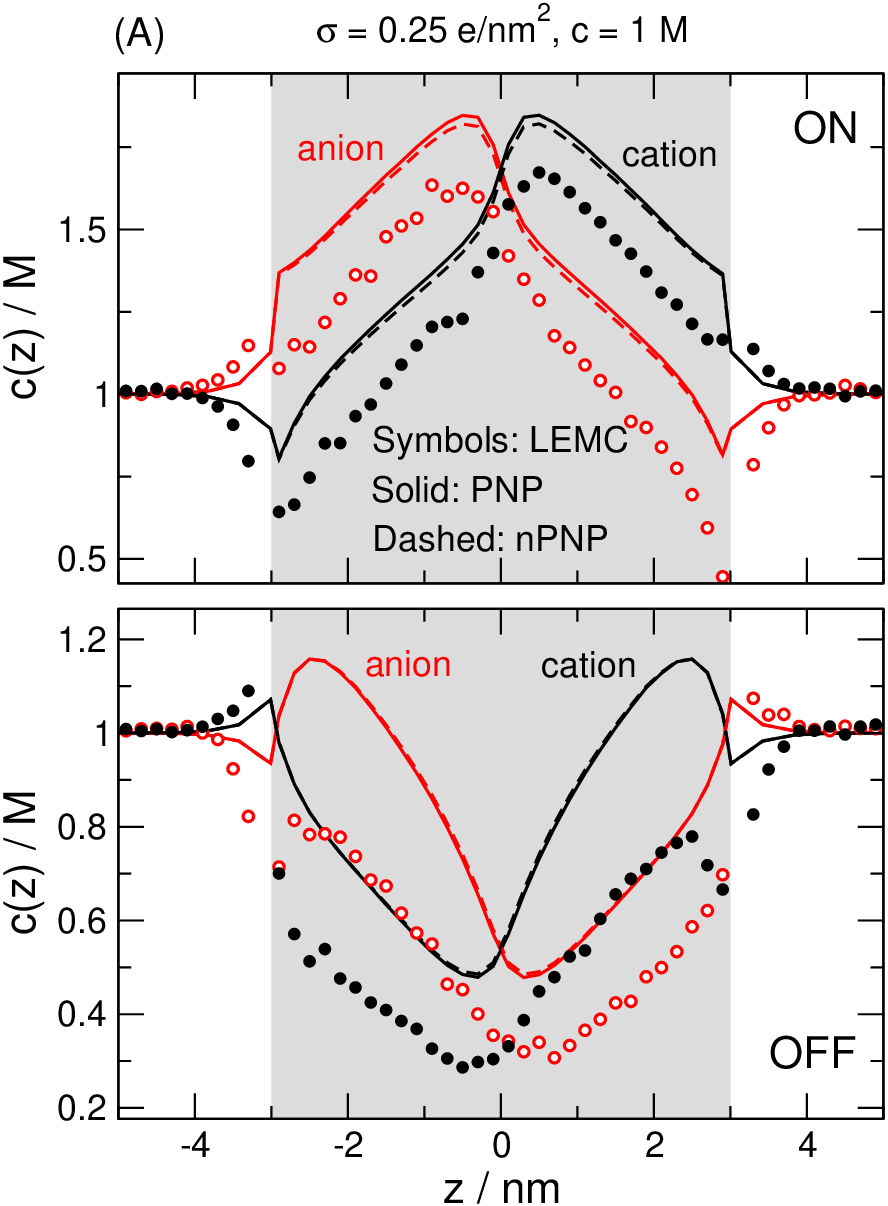}}\scalebox{0.493}{\includegraphics*{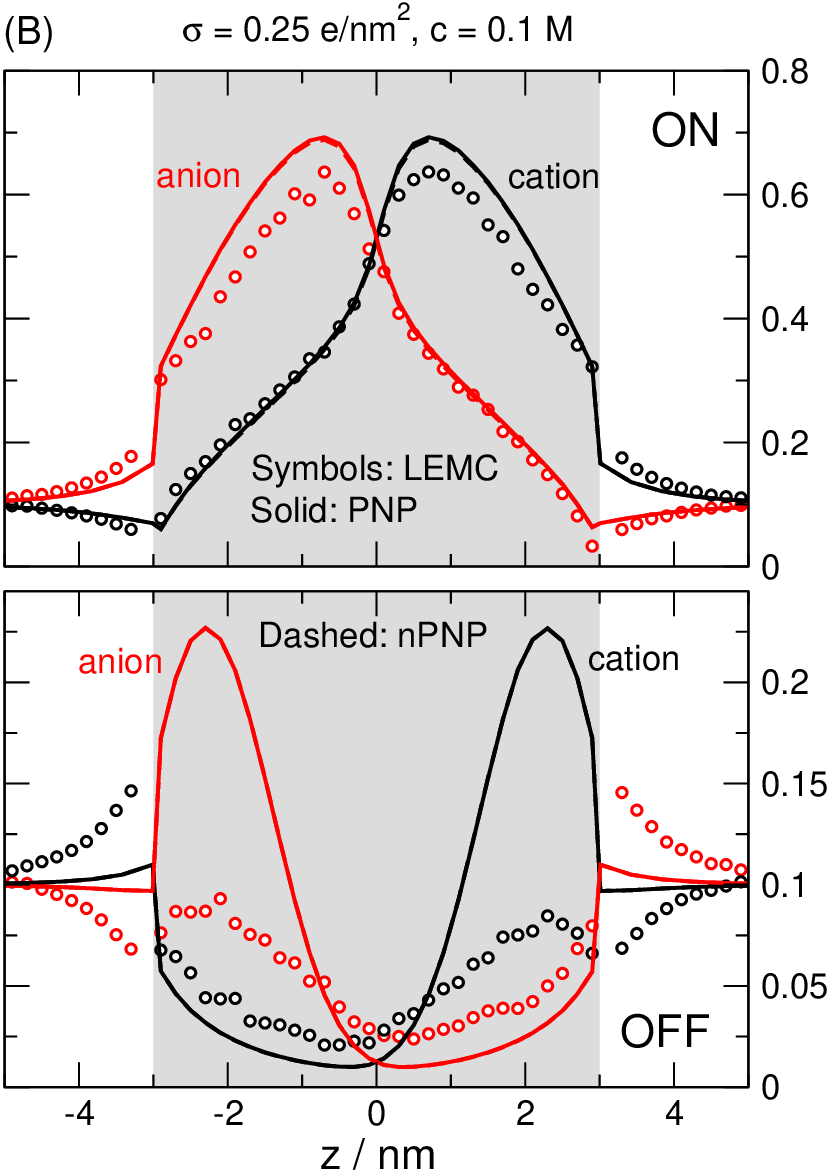}}
 \end{center}
\caption{\small 
Concentration profiles of cations and anions as obtained from NP+LEMC, PNP, and nPNP for (A) $c=1$ M and  (B) $c=0.1$ M for parameters $R=1$ and $\sigma=0.25$ $e$/nm$^{2}$.
}
\label{Fig7}
\end{figure*}

In Fig.\ \ref{Fig6}, we plot the concentration profiles for $c=1$ (panel A) and $0.1$ M (panel B) in order to study the differences between high and low concentrations.
This figure shows the results for $\sigma =1$ $e$/nm$^{2}$.
Figure \ref{Fig7} shows the same concentration profiles but for $\sigma=0.25$ $e$/nm$^{2}$.

The curves show that the ions have depletion zones in the middle of the pore and in the zone, where they are the co-ions (having ionic charge with the same sign as the pore charge, $\sigma$).
We distinguish basically four regions: 
\begin{enumerate}
 \item left bath, near the membrane ($z<-3$ nm)
 \item the left part of the pore with positive surface charge ($-3<z<0$ nm, N region) -- anions the counter-ions, cations the co-ions
 \item the right part of the pore with negative surface charge ($0<z<3$ nm, P region) -- cations the counter-ions, anions the co-ions
 \item right bath, near the membrane ($z>3$ nm)
\end{enumerate} 
In the access regions, close to the pore entrances (regions 1 and 4) ionic double layers are formed. 
Double layer is common name for the separation of cations and anions (polarization of the ionic distributions) as a response to the presence of a charged or polarized object.
In this case, double layers appear partly as a response to the applied field, partly as a response to the charge imbalance inside the pore.
Realize that the sign of the double layer (which ions are the co-ions and counter-ions in the double layer) depends on the sign of the applied voltage.

The basic reason of rectification is that the ions are more depleted in their depletion zones in the OFF state than in the ON state; cation concentration in the N zone is lower in the OFF state than in the ON state, for example.
Basically, the depletion zones are caused by the pore charges.
The applied field modulates the effect of pore charges, therefore, it increases or decreases concentrations compared to the zero-voltage case.
Depletion zones are the main determinants of the current, because they are the high-resistance elements of the system modeled as resistors connected in series along the ionic pathway.
So, if depletion zones are more depleted, current is reduced.

It is important, however, that not only the co-ion concentrations decrease by switching from ON to OFF, but also the counter-ion concentrations.
As a matter of fact, this is crucial, because co-ions are brought into their depletion zones with the help of their strong correlations to counter-ions. 
So there are less co-ions because there are less counter-ions.
The quantity of counter-ions, on the other hand, seems to be related to the double layers at the entrances of the pore on the two sides of the membrane.
At least, this seems to be suggested by the results of NP+LEMC.

\begin{figure*}[t]
 \begin{center}
\scalebox{0.48}{\includegraphics*{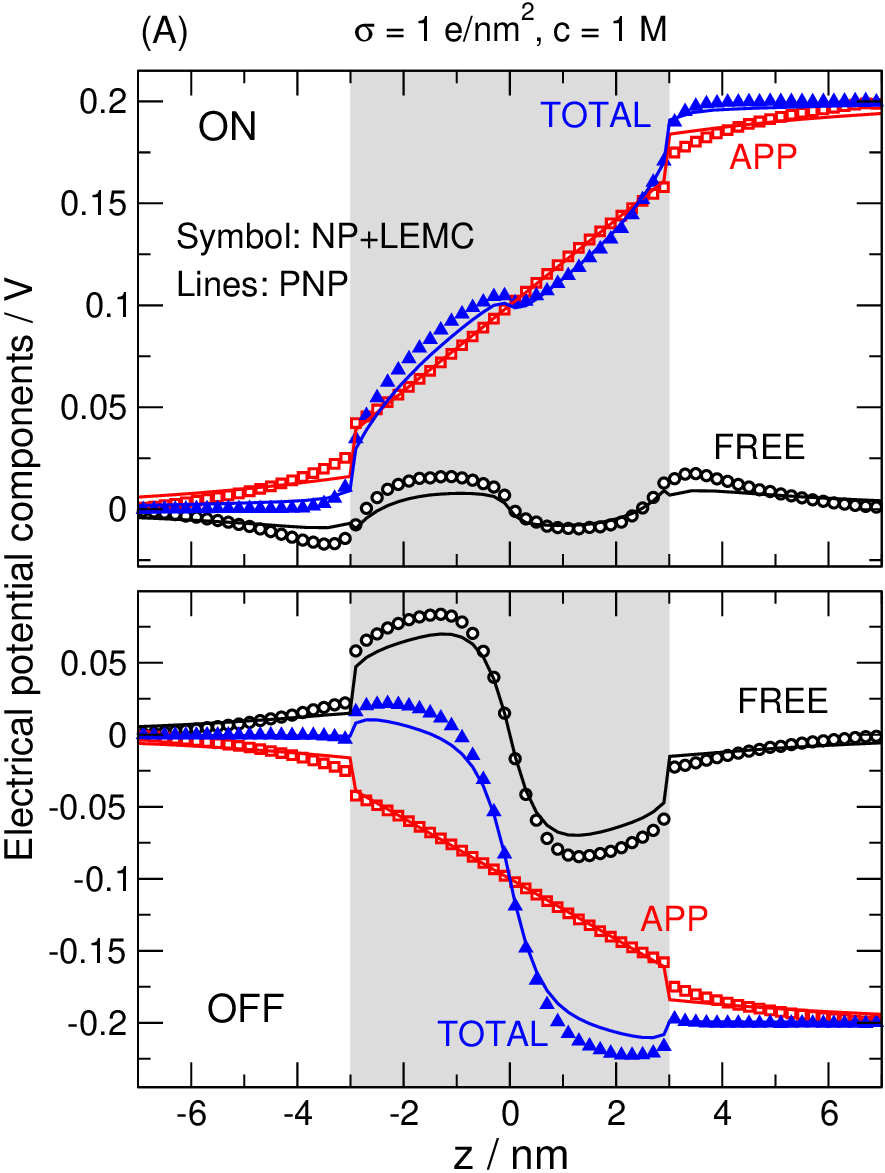}}
\scalebox{0.48}{\includegraphics*{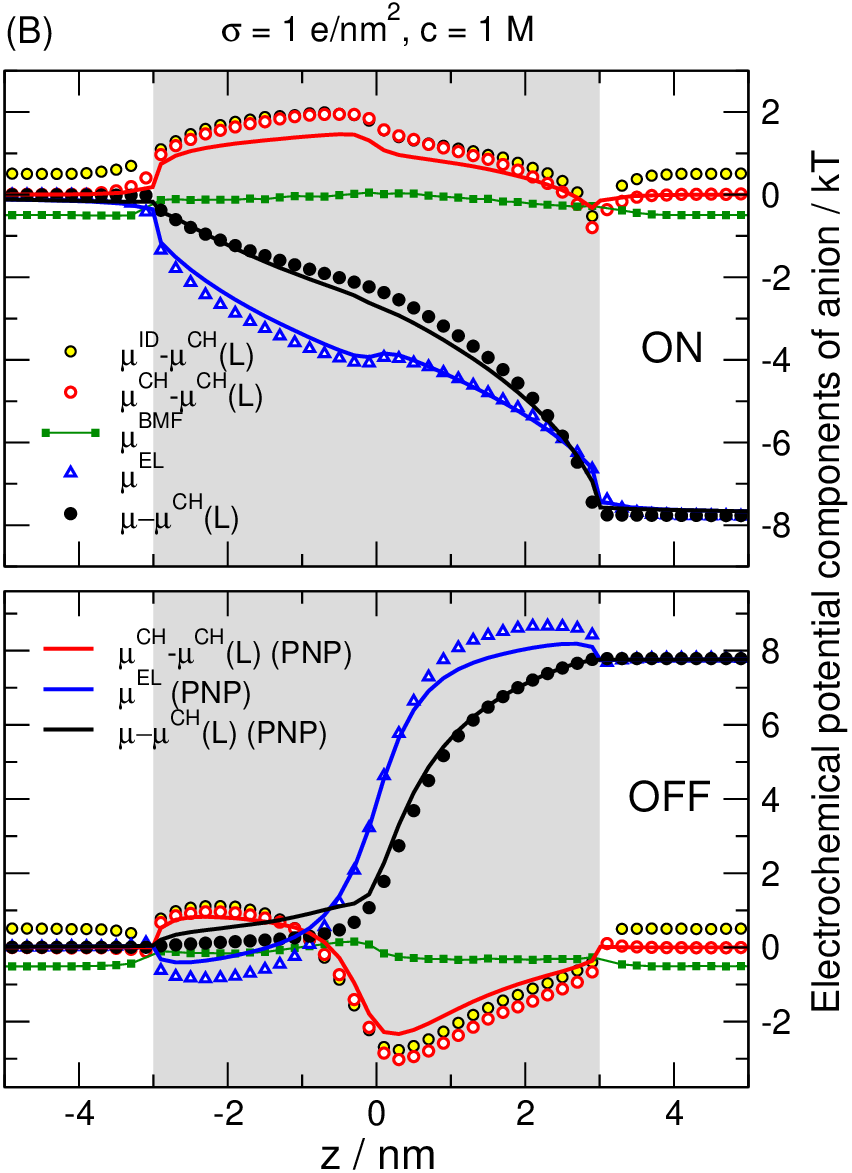}}
 \end{center}
\caption{\small (A) Electrical potential profiles and components (see Eq.\ \ref{eq:phi}) as obtained from NP+LEMC and PNP. Component $\Phi^{\mathrm{FREE}}(z)$ is the product of ions and pore charges in the system, while $\Phi^{\mathrm{APP}}(z)$ is the applied potential computed from the Laplace equation with Dirichlet boundary conditions.
(B) Electrochemical potential profiles and components (see Eqs.\ \ref{eq:elchempot}, \ref{eq:elchempotpnp}-\ref{eq:ch-plus-el}) as obtained from NP+LEMC and PNP.
The ideal ($\mu_{i}^{\mathrm{ID}}(z)=\mu_{i}^{0}+kT\ln c_{i}(z)$), the electrochemical ($\mu_{i}(z)$), and the chemical ($\mu_{i}^{\mathrm{CH}}(z)$) terms are shifted to zero by deducting $\mu_{i}^{\mathrm{CH}}(\mathrm{L})$, which is the value of the chemical term in the left bath.
In the case of PNP the ID and CH terms are the same, so $\mu_{i}^{\mathrm{BMF}}=0$.
Results are shown for the anion; data for the cation do not reveal new insights (subscript $i$ is dropped in the legend). 
Parameters are $c=1$ M, $\sigma=1$ $e$/nm$^{2}$, and $R=1$ nm. }
\label{Fig8}
\end{figure*}

The double layers have opposite signs in the ON and the OFF states that can be explained through the mean electrical potential profiles that have two components produced by all the free charges, $\Phi^{\mathrm{FREE}}(\mathbf{r})$, and induced charges, $\Phi^{\mathrm{APP}}(\mathbf{r})$,  in the system.
In this study, induced charges appear at the boundaries where the boundary conditions are applied, therefore, they produce the applied potential, $\Phi^{\mathrm{APP}}(\mathbf{r})$. 
The total mean potential, therefore, is obtained as
\begin{equation}
\Phi(\mathbf{r}) = \Phi^{\mathrm{FREE}}(\mathbf{r}) + \Phi^{\mathrm{APP}}(\mathbf{r}).
\label{eq:phi}
\end{equation}   
In the case of NP+LEMC, the double layers are necessary to produce the $\Phi^{\mathrm{FREE}}(z)$ component that counteracts the applied field, $\Phi^{\mathrm{APP}}(z)$.
Figure \ref{Fig8}A shows that the slope of $\Phi^{\mathrm{FREE}}(z)$ is the opposite to the slope of $\Phi^{\mathrm{APP}}(z)$ in the bulks, so their sum (TOTAL) has the slope close to zero.
This is necessary because the bulks are low-resistance elements, where the potential drop is small.

In the case of (n)PNP, this phenomenon depends on the imposed boundary conditions, Eq.\ \ref{eq:additionalneumann}, on the membrane surface.
Using, for example, $g=0$ yields totally different results which are in poor agreement with NP+LEMC as far as the structure of these double layers is concerned (the behavior inside the pore is less influenced).

Comparing the counter-ion profiles in the double layers and in the neighbouring half nanopores (Figs.\ \ref{Fig6} and \ref{Fig7}), we can see that if there are less counter-ions in the double layer, there are less counter-ions in the half nanopore too  (see anions on the left hand side in the OFF state compared to the ON state, for example).
Although the decrease of counter-ion concentration in the pore is related to the decrease of the concentration of the same ion in the neighbouring double layer, it would be an overstatement to say that one is a consequence of the other.  

Rectification works without this coupling between ion quantities in the double layer and in the nanopore.
For example, rectification is reproduced in the case of PNP with boundary condition $g=0$ although with worse agreement with NP+LEMC. 
Furthermore, the formation of the double layers is absent in MD simulations using explicit water, still, rectification is present.
MD results using explicit water are in good agreement with NP+LEMC results using implicit water \cite{hato_nl_2017}.
These contradictions require more study, but it seems that the formation of the double layers is rather related to boundary conditions and larger-scale effects, while the structure of the ionic profiles inside the pore is rather related to local effects such as interaction with pore charges, applied field, and other ions. 

As far as the agreement between the NP+LEMC and the theoretical profiles is concerned, it is generally better in the ON state than in the OFF state (see Figs.\ \ref{Fig6} and \ref{Fig7}).
In the OFF state, (n)PNP usually overestimates concentrations causing the overestimation of current as we have seen before.
This is counterintuitive, because it was said that (n)PNP is better at low concentrations, but pore concentrations are higher in the ON state.
We can resolve this contradiction if we consider that the system's behavior is a result of the balance of basically three effects: (1) interaction with the fixed pore charges, (2) interaction with the fixed applied field, and (3) mutual and complicated interactions between ions. 
The mutual weight of these terms is different in the ON and OFF states.

In the ON state, pore charges and applied field act in the same direction, so they dominate the energy and errors in the ion-ion term have less effect.
In the OFF state, however, pore charges and applied field act in the opposite directions, so their sum is smaller and the ion-ion term has a larger weight and the BMF term with it.

Our next goal is to better understand the different contributions of the components of the total electrochemical potential $\mu_i$ as defined in Eqs.\ \ref{eq:elchempot}--\ref{eq:ch-plus-el}. 
The electrical component $\mu_i^{\mathrm{EL}}$ is defined as the interaction with the (total) mean electrical potential that is shown in Fig.\ \ref{Fig8}A. 
Note that the BMF term is fully included in the CH term and therefore, in the case of (n)PNP, $\mu_i^{\mathrm{CH}}(\mathbf{r})$ is just $\mu_{i}^{\mathrm{ID}}(\mathbf{r})$, while it also contains the BMF term in the case of NP+LEMC.

Figure \ref{Fig8}B shows the full electrochemical potentials, the CH terms, and the EL terms.
In the case of NP+LEMC, we also plot the $\ln c_{i}(z)$ term (denoted as ID) and the BMF term.
The ID and CH terms, as well as the total electrochemical potential, are all shifted by the value of the CH term in the left bath ($\mu_{i}^{\mathrm{CH}}(\mathrm{L})$).
In this way, the $\mu_{i}(z)$, $\mu_{i}^{\mathrm{CH}}(z)$, and $\mu_{i}^{\mathrm{EL}}(z)$ contributions take the value zero at the left edge of the plot.

The errors in $\mu_i$ have three components: the error in reproducing (1) the $\ln c_{i}$ term, (2) the EL term, and (3) the BMF term that can be identified with errors in reproducing the particle correlations which are missing in PNP, due to the mean field approximation.
The first two errors have different signs and tend to balance each other. 
They are coupled through the Poisson equation, so in the limiting case of agreeing $c_{i}$ profiles, the $\Phi$ profiles agree if the boundary conditions are also the same.

In this case, the NP equation would give the same flux if the BMF term were constant, because $\nabla \mu_{i}$ would be the same in the two methods.
Therefore, the real source of errors is not the magnitude of the BMF term, but the $\mathbf{r}$-dependence of the BMF term, that is, the fact that ionic correlations are different inside the pore than outside.
The nonzero value of the BMF term, on the other hand, indicates that there is an error in ``chemistry'', so there is a possibility for further errors in both the $c_{i}(\mathbf{r})$ and $\Phi(\mathbf{r})$ profiles inside the pore.
Those potential errors can eventuate inside the pore and become visible in all profiles.
Local fluctuations in the BMF term inside the pore indicate how seriously do the errors of the mean-field treatment of PNP contribute to inaccuracies of all the profiles inside the pore.

\section{Summary}
\label{sec:summary}

The general conclusion is that the BMF term is small and the agreement between PNP and NP+LEMC is quite good. 
Yet, since the mean field theory does not capture the OFF state behavior as good as the ON state behavior, derived quantifies as the rectification cannot be predicted that well.
Still, the results are very promising given that these calculations have been performed for a narrow ($R = 1$ nm) and short ($6$ nm) pore with experimentally typical, but quite large surface charges ($\sigma \sim 1$ $e$/nm$^{2}$). 
This indicates that the 2D PNP used in this study is an appropriate tool to study more realistic geometries (wider and longer pores), at least, as far as the agreement with simulations in the framework of an implicit solvent model is concerned.

This work is a link in a series of works, where a given system (a bipolar nanopore) is studied using different levels of modeling. 
Our results only prove that PNP calculations are useful in the framework of an implicit water model.
Whether the implicit water model is a useful one is the topic of another publication \cite{hato_nl_2017}, where we compare implicit-water NP+LEMC simulations with explicit-water MD simulations.

A real nanopore is obviously too big to use MD simulations and all-atom models as a general tool.
Although computers are getting faster and faster, the quality of force fields seems to be a serious limiting factor.
Still, all-atom (in this case, this means explicit water) MD simulations can be done for the nanopore of the size studied in this work. 
Therefore, MD simulations can have a crucial role in a chain of calculations, where we increase the complexity of modeling step by step.

In general, particle simulation studies are more useful where local effects are important.
The typical example is the narrow bottleneck of a nanopore, where double layers overlap.
Nanopores can also be used as sensors \cite{sexton_mbs_2007,howorka_csr_2009,vlassiouk_jacs_2009,piruska_csr_2010,ai_sab_2011,howorka_nbt_2012}, where the detectable analyte molecule is selectively bound by a binding site of another molecule that is attached to the tip of the nanopore.
The binding of the analyte molecule influences the effective cross section, and, thus, the current.
An associated and thoroughly studied phenomenon is the crossing the a DNA molecule through the nanopore during which the sequencing might be possible in an efficient and fast manner \cite{peng_chap11_2011,otto_chapter_2013}. 
These are obviously local effects, where particle simulations are useful.

The device itself that is around the tip of the nanopore, however, is too big to compute with particle simulations using its real dimensions. 
In general, it is our purpose to model phenomena with their appropriate boundary conditions using close to real time and length scales at least on the mesoscopic level.
This purpose can be achieved using the multiscale modeling framework in which the advantages of all the modeling levels and associated computation methods can be exploited.

This series of calculations proves that reducing the models by neglecting certain effects is an appropriate procedure for the case of ionic solutions and the bipolar nanopore studied here. 
This is also due to the fact that the transport of ions is mainly determined by electrostatic effects.
The interactions with the applied field, permanent surface charges, and other ions treated on a mean field level are sufficient to reproduce the system's basic behavior. 
For different systems, procedures similar to this should be repeated in order to evaluate the validity of the mean field approximation.

\section*{Acknowledgments}

DB and MV gratefully acknowledge the financial support of the Hungarian National Research Fund (OTKA NN113527) in the framework of ERA Chemistry. 
The work of JFP was supported by DFG via Grant 1073/1-2. 
MTW and BM acknowledges financial support from the Austrian Academy of Sciences \"OAW via the New Frontiers Grant NST-001.

%

\end{document}